\documentclass[prl,aps,twocolumn,showpacs,groupedaddress,amsmath,amssymb,floatfix]{revtex4}
\usepackage{graphicx}

\newcommand{\be}{\begin{equation}}
\newcommand{\ee}{\end{equation}}
\newcommand{\xx}{{\mathbf x}}

\newcommand{\kk}{{\mathbf k}}
\newcommand{\qq}{{\mathbf q}}

\newcommand{\eq}[1]{(\ref{#1})}
\newcommand{\mume}{\mu \rm m}

\begin{document}
\title{Creation and detection of vortices in polariton condensates}
\author{Michiel Wouters}
\affiliation{Insitute of Theoretical Physics, Ecole Polytechnique F\'ed\'erale de Lausanne (EPFL), CH-1015 Lausanne, Switzerland}
\author{Vincenzo Savona}
\affiliation{Insitute of Theoretical Physics, Ecole Polytechnique F\'ed\'erale de Lausanne (EPFL), CH-1015 Lausanne, Switzerland}
\begin{abstract}
We investigate theoretically the creation, persistence and detection of quantized vortices in nonequilibrium polariton condensates within a stochastic classical field model. The life time of the quantized vortices is shown to increase with the spatial coherence and to depend on the geometry of the condensate. The relation with superfluidity in conventional superfluids is discussed. Three different ways to measure the vorticity of the polariton condensate are proposed.
\end{abstract}
\maketitle


The connection between the two intriguing phenomena of superfluidity and long range spatial coherence was conjectured soon after the discovery of superfluidity. It has proven to be subtle, but is now well established for Bose-Einstein condensates at thermal equilibrium~\cite{leggett_book}. 
The recent experimental observation of the spontaneous build up of long range spatial coherence in exciton-polariton gases~\cite{kasprzak} as well as a bimodal momentum distribution~\cite{kasprzak,snoke,yamamoto} and long temporal coherence~\cite{love2008}, has induced the interest in the phenomenon of Bose Einstein condensation out of thermal equilibrium. Due to the short (ps) life time of the polaritons, they have to be continuously injected in the microcavity by means of relaxation from an excitonic reservoir that is excited by a laser. The resulting steady state is not a thermal equilibrium one, yet it still features the main characteristic of an equilibrium Bose-Einstein phase transition: short spatial coherence below a critical density, long range coherence above. A natural question is whether the phase with long range coherence manifests superfluid properties. The identification of a superfluid regime out of equilibrium has been addressed for a resonantly driven polariton gas, where the coherence is imprinted by the excitation laser~\cite{iac-superfl,amo-superfl}. Under this excitation condition, superfluidity manifests itself as a reduction in the Rayleigh scattering when the polariton gas is injected at finite momentum. In this Letter, we will address the issue of superfluidity of a nonresonantly excited polariton gas where the long range order is spontaneously formed and superflows are not directly forced by the excitation laser.

Phenomenologically, superfluidity can be loosely described as `flow without friction', referring to the remarkable property that the superfluid does not necessarily move along with the container that holds it. This idea of `flow without friction' does not apply to driven-dissipative systems: When energy is fed into the system, it does not come as a surprise that the fluid is in motion with respect to its container. This is why it was not possible to interpret the observation of the quantized vortices in polariton condensates~\cite{konstantinos} as a proof of superfluidity. The appearance of vortices in the steady state of a polariton condensate was rather interpreted as a result of the flow driven by the relaxation of high energy excitons into the lower polariton branch that is subject to an irregular potential, a phenomenon that could equally well be exhibited by a classical system. The same holds for the spontaneously formed vortex lattice that was predicted in the theoretical work of Ref.~\cite{keeling}. 

A slightly more general view on a superfluid is that it supports many metastable flow patterns under the same boundary conditions. Such behavior is not displayed by classical systems: the velocity of water in a pipe with a given pressure difference between both ends converges quickly to its unique steady state. This definition of superfluidity as multiple metastable flow patterns is equally meaningful in nonequilibrium systems. We will propose an experiment to investigate the superfluidity of polariton condensates in this sense and make its theoretical analysis within the frame work of the classical field model introduced in Ref.~\cite{fluctuations}. In particular, we will investigate the evolution of an additional vortex introduced in a vortex free polariton condensate. 

A mechanical creation of vortices, in analogy with superfluid $^4$He and ultracold atomic gases~\cite{abo,madison} is probably not practical, because it would require mechanical rotation frequencies between 1 GHz and 1 THz, but
thanks to the half light nature of polaritons, a vortex can be created with a pulsed~\footnote{Our proposal differs in this respect from the experiments reported in Ref.~\cite{dimavortex}, where a vortex was created with a cw resonant laser.} resonant laser beam that contains one unit of angular momentum per photon. 
Because it is at present technologically impossible to perform single shot measurements, direct visualization of a moving vortex in the density or phase profile of the condensate cannot be achieved. We will propose instead three different operational methods to detect the condensate vorticity in multiple shot measurements.

Our classical field model describes the coupled dynamics of a classical incoherent exciton reservoir to the quantum dynamics in the lower polariton branch. The coupled equations of motion of the reservoirs describe the nonresonant excitation with intensity $P$, and the scattering into the lower polariton region
\begin{multline}
\frac{d}{dt} \tilde n_{R}=P-\gamma_{R} \tilde n_{R}- \alpha\frac{d}{dt}\langle \psi^\dag \psi\rangle |_{\rm res}  \\
-\frac{\alpha}{2\Delta V}\sum_\kk 
[\tilde n_R^2 R_{\rm in}[\epsilon_{LP}(\epsilon(\kk)]- \tilde n_R R_{\rm out}[\epsilon(\kk)].
\label{eq:motres}
\end{multline}
The coupling terms between the reservoir and lower polariton branch read explicitely~\footnote{The geometric mean of $n_{R2}$ at positions $\xx$ and $\xx'$ is a slight modification with respect to the one that is derived from a microscopic Hamiltonian, that simplifies the generation of the noise.}
\begin{multline}
	\mathcal R_{\rm in, out} \psi(\xx) =  \sum_{\qq,\xx'} e^{i\qq (\xx-\xx')} \sqrt{n_{R2}(\xx)n_{R2}(\xx')}\\
	\times R_{\rm in,out}(\epsilon_\qq)  \psi(\xx')
	\label{eq:kernel}
\end{multline}
The reservoir density $\tilde n_R$ is normalized so that threshold is reached when it is unity. The lower polariton dynamics is described by the stochastic classical field equation
\begin{multline}
i d \psi = dt\; \left[\frac{-\hbar^2 \nabla^2}{2m}+ \frac{i(\mathcal R_{\rm in}-\mathcal R_{\rm out}-\gamma)}{2}+ 
\frac{g}{\Delta V} |\psi(\xx)|^2 \right]\psi \\+ dW.
\label{eq:stoch_mot}
\end{multline}
The noise term $dW$ is a complex Gaussian stochastic variable with the correlation functions:
\begin{multline}
\langle dW(\xx) dW (\xx') \rangle =0, \hspace{1cm}\\
\langle dW(\xx) dW^* (\xx') \rangle =
 \frac{dt}{4 \Delta V} \left(\langle \xx| \mathcal R^S_{\rm  in}+\mathcal R^S_{\rm  out}| \xx' \rangle +2 \gamma \delta_{\xx,\xx'}\right),
\label{eq:noisecor}
\end{multline}
where $\mathcal R^S_{\rm in,out}=[\mathcal R_{\rm in,out}+(\mathcal R_{\rm in,out})^{\rm T}]/2$ are the symmetrized in and out-scattering kernels. 
According to Eq.~\eq{eq:stoch_mot}, the number of excitons that scatter from the reservoir into the lower polariton classical field equals $\frac{d}{dt}\langle \psi^\dag \psi\rangle |_{\rm res} = 2 {\rm Re}[\psi^*(\mathcal R_{\rm in}-\mathcal R_{\rm out})\psi]$. The parameter $\alpha$ in Eq.~\eq{eq:motres} quantifies the strength of the  coupled to the lower polariton field $\psi$; the last term in Eq.~\eq{eq:motres} comes from the fact that the classical field in Eq.~\eq{eq:stoch_mot} contains 1/2 zero point fluctuations per mode, because it samples the the Wigner quasi-probability distribution $P_W$, that is related to the observable density as $\int P_W(\psi) \psi^*\psi=\langle \hat \psi^\dag \hat \psi+\hat \psi \hat \psi^\dag \rangle/2$. In the simulations, we have used the following parameters:
$\gamma=0.5 \; {\rm meV}$, $m_{LP}/\hbar = 1  \;\mume^{-2} {\rm meV}^{-1}$, $g/\hbar=0.005 \; {\rm meV}  \mume^{2}$,  $R_{\rm out}(E)=(3 {\rm meV})$, $R_{\rm in}(E)=\exp(-E/k_B T_R)[\gamma+3 {\rm meV}]$, with $k_B T_R=2\; {\rm meV}$. Numerical results are obtained by averaging over 100 realizations of the noise.

\begin{figure}[htbp]
\begin{center}
\includegraphics[width=\columnwidth,angle=0,clip]{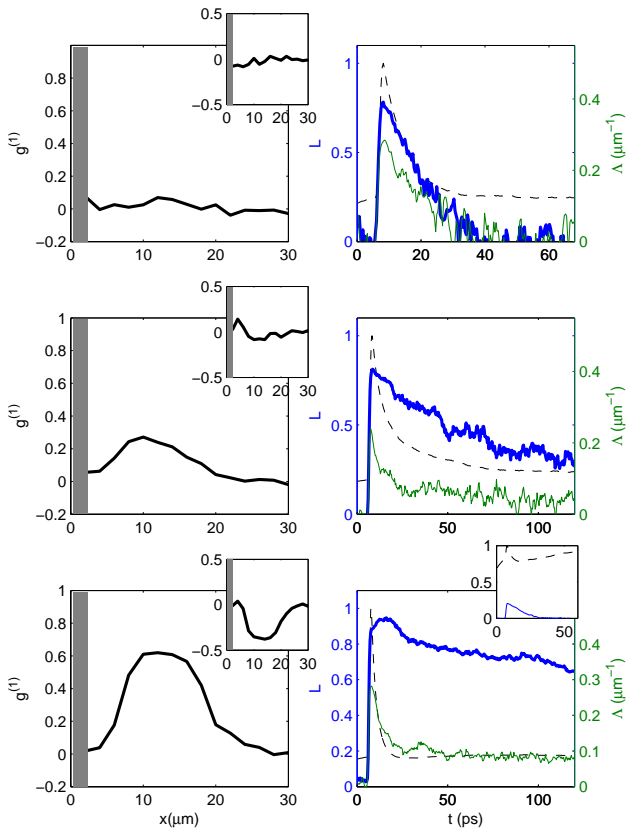}
\end{center}
\caption{The left column show the first order spatial coherence between inversion symmetric points $g^{(1)}(x)=\langle \psi^\dag(-\xx) \psi(\xx) \rangle /\sqrt{n(\xx)n(-\xx)}$ for values of the density increasing from left to right ($n=1.3, 5.7\;{\rm and } \;32\; \mume^{-2}$ respectively), before the arrival of the Gauss-Laguerre pulse; the insets show the values at the latest simulated times. The right column shows the temporal evolution of the total polariton density (dashed line, normalized to its maximal value) and angular momentum (thick) after a Gauss-Laguerre probe pulse is has perturbed the polariton condensates. The thin line shows the observable $\Lambda$ that we propose to use for a time resolved measurement of the vorticity. The inset in the lower right figure shows the response of the condensate to a probe pulse that is too weak to create a vortex. The polariton condensate is created with a ring shaped excitation laser (inner radius 7$\mume$; outer radius 20$\mume$).}
\label{fig:acrossthr1}
\end{figure}

Vortices are typically more stable in a toroidal geometry as compared to a simply connected one. We therefore start the analysis of the vortex life time for a polariton gas that is excited by a ring shaped excitation laser. The right hand panels in Fig.~\ref{fig:acrossthr1} show the simulated evolution of the angular momentum and excess density when a Gauss-Laguerre beam with one unit of angular momentum is applied to the polariton gas for three different condensate densities (increasing from top to bottom). The corresponding spatial coherence is shown in the left hand panels. The excess density (dashed line) evolves almost identically in the three cases~\footnote{Note however the slower decay of the excess density when closer to threshold, related to the critical slowing down, see Ref.~\cite{ballarini} for an experimental and theoretical discussion in the parametric oscillation regime.}, but the time dependence of the angular momentum is dramatically different. At low densities, the angular momentum quickly decays together with the excess density introduced by the probe pulse, where at the highest density, the angular momentum lasts up to the end of our simulation time window (120 ps). The insets in the left column show the spatial coherence at the final time of our simulation window. In the upper and central panels, where the vortex disappears at the final time of the simulations, an important decrease in the absolute value of the coherence is observed.

\begin{figure}[htbp]
\begin{center}
\includegraphics[width=\columnwidth,angle=0,clip]{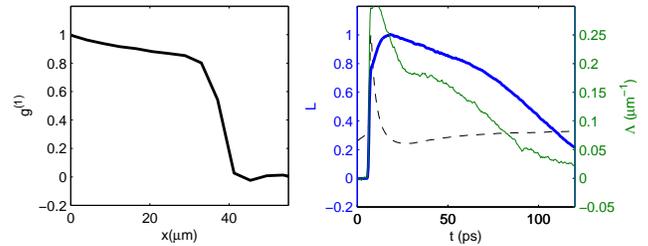}
\end{center}
\caption{The same as Fig. 1, but for a condensate excited with a top-hat excitation laser with a 20$\mume$ radius (maximum density $n=48\; \mume^{-2}$).}
\label{fig:acrossthr2}
\end{figure}

The simultaneous increase in the coherence and the vortex life time is in analogy with conventional superfluids. The argument for the difference in life time of the angular momentum is that when the phase of the condensate is coherent over large distances, a local distortion of the phase is not possible and vorticity can only disappear if the vortex core moves out of the polariton gas. The time it takes for a vortex to move out depends on the geometry of the condensate. In polariton condensates, an important phenomenon that determines the life time of the superflow are the currents that arise due to the interplay between the nonresonant excitation and the external potential~\cite{woutersprb08}. Vortices are dragged by these currents and in order to achieve a long life time of the vortex it is beneficial if these currents are pointed toward the center of the vortex. Under excitation with a ring shaped excitation laser of Fig.~\ref{fig:acrossthr1}, the blue shift causes the polaritons to flow toward the center, which hinders the vortex to leave the system, because it has to move upstream. Fig.~\ref{fig:acrossthr2} shows the same simulations with a regular top hat excitation laser. The polariton-polariton interactions now cause an outward instead of inward flow and the angular momentum in the phase coherent regime decays much quicker. 

The phase rigidity of the polariton condensate becomes also clear in the response of the angular momentum for varying intensity of the Gauss-Laguerre pulse. At low intensities, the pulse is not able to create a vortex in the condensate so that the transfer of angular momentum is poor and decays quickly. Only when a certain threshold intensity is passed (corresponding in our simulations to an excess density of about half the condensate density) a vortex is formed and the angular momentum persists for long times. This behavior is illustrated in the inset in the lower right panel of Fig.~\ref{fig:acrossthr1}, where the response to a low intensity pulse is shown. In the normal, incoherent phase, we have observed no qualitative change in the transfer of angular momentum: it always decays on the same time scale as the excess density introduced by the pulse.

We now turn to the question how the metastable superflow can be detected experimentally. Ideally, an experimental measurement should be time resolved, so to observe the predicted increase in life time of the metastable superflow when the condensate density is increased.

A first technique is the one that was used previously to detect vortices induced by the disordered potential acting on the polaritons: A vortex leads to a fork dislocation in the interference pattern of the condensate with its inverted copy~\cite{konstantinos}. In order to observe the increased life time of the superflow when the condensate density is increased,  this interferogram should be measured with time resolution. This is possible with the use of a streak camera. Still, an average over many runs of the experiment should be recorded in order to have a good signal to noise ratio. This poses some difficulties when the vortex undergoes a random motion that is different from shot to shot. In that case, no dislocation will be visible in an interferogram that is averaged over many realizations. 

A first possibility to do a time resolved measurement is based on the change of sign of $g^{(1)}(\xx,-\xx)$ (compare the left hand plots in Fig.~\ref{fig:acrossthr1} with their insets). In a time-resolved interferogram, such as the one recorded in Ref.~\cite{condform}, this would be visible as an interchange between light and dark fringes as a function of time. The random motion of the vortex leads to a reduced contrast of the fringes, but fringes are not washed out (the coherence between two points at opposite positions of the vortex in the right panel of Fig.~\ref{fig:acrossthr1} is about 0.4).

\begin{figure}[htbp]
\begin{center}
\includegraphics[width=\columnwidth,angle=0,clip]{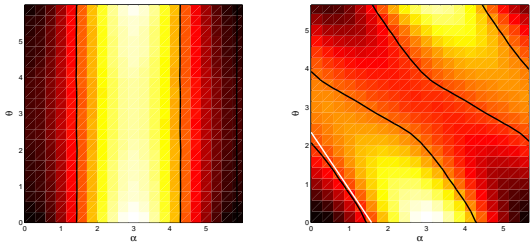}
\end{center}
\caption{Interferogram as a function of the phase difference $\alpha$ and rotation angle $\theta$ betweeen the copies of the condensate that are interfered,  before (t=0 ps, left hand panel) and after (t=100 ps, right hand panel) the arrival of the Gauss-Laguerre pulse. The black lines show the contour lines $I(\alpha,\theta)=1$; the white line shows $\alpha=\pi/2+\langle \hat L_z \rangle \theta$, where $\langle \hat L_z \rangle$ is taken from Fig.~\ref{fig:acrossthr1}. The presence of angular momentum is signalled by the tilting of the interference fringes in the $(\alpha,\theta)$ plane.}
\label{fig:interfero}
\end{figure}

A second possibility is a direct measurement of the time dependence of the angular momentum. We discuss here how this can be achieved with interferometry. The operator that rotates a state over an angle $\theta$ can be written as $\hat R_{\theta}=\exp(i \theta \hat L_z)$. Letting the polariton condensate interfere with a copy that is rotated by an angle $\theta$ and phase shifted by $\alpha$ leads to a integrated intensity, normalized to the total number of particles $N$
\begin{eqnarray}
I(\alpha,\theta) &=&\frac{1}{2 N}\int d\xx \langle | \hat \psi(\xx) + e^{i\alpha}   \hat \psi(R_{-\theta} \xx) |  ^2 \rangle  
\\
 &=& 1 - \frac{i \hbar}{N}\int d\xx \langle \hat \psi^\dag(\xx) \cos[\theta \, (\xx \times \nabla)_z + \alpha] \hat \psi(\xx)  \rangle   \notag,
\end{eqnarray}
For small rotation angles and $\alpha\approx \pi/4$, the argument of the cosine can be expanded. It is then easy to see that the contour line of the interferogram $I(\alpha,\theta)=1$ goes to $\langle \hat L_z \rangle \theta=\pi/2-\alpha$. If the fluctuations of the angular momentum are small, the contour lines of the interferogram are straight lines and the angular momentum can be determined from a large region in the $(\alpha,\theta)$ plane.
Fig.~\ref{fig:interfero} shows simulated interferograms. Left and right panels are before and after the Gauss-Laguerre pulse respectively. In the right hand panel, the vorticity of the condensate results in a tilting of the interferogram in the $(\alpha,\theta)$ plane. Note the decrease in the contrast of the interferogram at finite rotation angles $\theta$. This is a consequence of the imperfect spatial coherence.

\begin{figure}[htbp]
\begin{center}
\includegraphics[width=0.8\columnwidth,angle=0,clip]{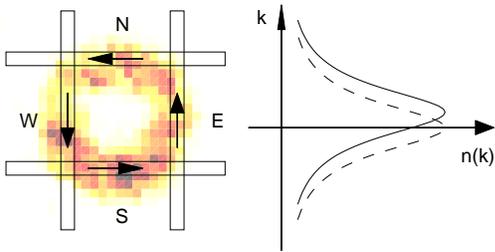}
\end{center}
\caption{Sketch of the scheme to detect the extra velocity in the polariton condensate due to the presence of a vortex: The momentum distribution from a slice selected in real space is shifted in the presence of a vortex (full line) with respect to the momentum distribution in its absence (dashed line).}
\label{fig:sketch}
\end{figure}

An alternative scheme that could be used to detect the presence of a vortex is based on the fact that a vortex induces a flow of polaritons $v\propto {\bf 1}_\theta/r$. Therefore, measuring the velocity (momentum) of polaritons on a path around the vortex can reveal its presence. The proposal to measure this velocity is sketched in Fig.~\ref{fig:sketch}. It consists of measuring the momentum distribution of a slice of polaritons selected in real space. The presence of vortex will be signaled by a shift in the momentum distribution along the long directions of the slits. The observable that corresponds best to the angular momentum is 
$\Lambda = \langle k_{E} - k_{N}- k_{W}+ k_{ S} \rangle$. It is plotted in Figs.~\ref{fig:acrossthr1} and~\ref{fig:acrossthr2} together with the angular momentum $L$, showing a good correspondence between $\Lambda$ and $L$. Note that this technique works better in the case of a regular top hat excitation laser, where the vortex sits in a high density region. In the ring excitation geometry, this detection technique is more difficult, because no region with simultaneously a sizable density and velocity can be found.

In conclusion, we have theoretically investigated the possibility to create and detect quantized vortices in polariton condensates by resonantly exciting them with a Gauss-Laguerre laser beam. Only when the system exhibits a large degree of long range order, the life time of the angular momentum substantially exceeds the life time of the excess density introduced by the Gauss-Laguerre beam. We have proposed and analyzed several schemes to experimentally detect the presence of a quantized vortex.

It is a pleasure to thank I. Carusotto, D. Sarchi and K. Lagoudakis for stimulating discussions. We acknowledge the support of NCCR Quantum Photonics (NCCR QP), research instrument of the Swiss National Science Foundation (SNSF).


\begin{thebibliography}{99}
\bibitem{leggett_book} A. J. Leggett, {\em Quantum Liquids: Bose Condensation and Cooper Pairing in Condensed-Matter Systems} (Oxford, 2006).


\bibitem{kasprzak} J. Kasprzak {\em et al.}, Nature {\bf 443}, 409 (2006).

\bibitem{yamamoto} H. Deng \emph{et al.}, Phys. Rev. Lett. \textbf{97}, 
146402 (2006).

\bibitem{snoke} R. Balili, V. Hartwell, D. Snoke, L. Pfeiffer, and K. West, Science {\bf 316}, 1007 (2007). 

\bibitem{love2008}  A. P. D. Love, D. N. Krizhanovskii, D. M. Whittaker, R. Bouchekioua, D. Sanvitto, S. Al Rizeiqi, R. Bradley, M. S. Skolnick, P. R. Eastham, R. Andr\'e, and Le Si Dang, Phys. Rev. Lett. { \bf 101}, 067404 (2008).

\bibitem{iac-superfl} I. Carusotto and C. Ciuti, Phys. Rev. Lett. {\bf 93}, 166401 (2004).

\bibitem{amo-superfl}  A. Amo {\em et al.}, arXiv:0903.2723.

\bibitem{konstantinos} K. G. Lagoudakis, M. Wouters, M. Richard, A. Baas, I. Carusotto, R. Andr\'e, Le Si Dang, B. Deveaud-Pl\'edran, Nature Physics {\bf 4}, 706 (2008).

\bibitem{keeling} J. Keeling and N. G. Berloff, Phys. Rev. Lett. {\bf 100}, 250401 (2008).

\bibitem{fluctuations} M. Wouters and V. Savona, ArXiv:0811.4567, Phys. Rev. B in press.

\bibitem{abo} J. R. Abo-Shaeer, C. Raman, and W. Ketterle, Phys. Rev. Lett. {\bf 88}, 070409 (2002).

\bibitem{madison} K. W. Madison, F. Chevy, W. Wohlleben, and J. Dalibard, Phys. Rev. Lett {\bf 84}, 806 (2000). 


\bibitem{woutersprb08} M. Wouters, I. Carusotto and C. Ciuti, Phys. Rev. B {\bf 77}, 115340 (2008).

\bibitem{condform} B. Pietka {\em et al.} in preparation.


\bibitem{dimavortex} D. N. Krizhanovskii {\em et al.}, in preparation.
	
\bibitem{ballarini} D. Ballarini, D. Sanvitto, A. Amo, L. Vi\~na, M. Wouters, I. Carusotto, A. Lemaitre, and J. Bloch,  Phys. Rev. Lett. {\bf 102}, 056402 (2009).



\end{thebibliography}
\end{document}